\documentclass[aps,twocolumn,amsfonts,amssymb,amsmath,showpacs]{revtex4}
\usepackage{epsfig, graphicx, color, amsmath, amsthm, amssymb}
\def\ket #1{\vert #1\rangle}
\def\bra #1{\langle #1\vert}

\def\abs #1{\lvert #1\rvert}
\DeclareMathOperator{\tr}{Tr}
\newcommand{\beq}{\begin{equation}}
\newcommand{\eeq}{\end{equation}}

\newlength{\commentslength}

\newtheoremstyle{note}{}{}{\slshape}{}{\bfseries}{.}{ }{}
\theoremstyle{definition}
\newtheorem{remark}{Remark}

\begin{document}

\author{Ben W. Reichardt}
\email{breic@cs.berkeley.edu}
\affiliation{{EECS} Department, Computer Science Division, University of California, Berkeley, California 94720}

\author{Lov K. Grover}
\email{lkgrover@bell-labs.com}
\affiliation{Bell Laboratories, Lucent Technologies, 600-700 Mountain Avenue, Murray Hill, New Jersey 07974}


\pacs{03.67.Pp, 82.56.Jn}

\title{Quantum error correction of systematic errors using a quantum search framework}

\begin{abstract}
Composite pulses are a quantum control technique for canceling out systematic control errors.  We present a new 
composite pulse sequence inspired by quantum search.  
Our technique can correct 
a wider variety of systematic errors -- including, for example, nonlinear over-rotational errors -- than previous techniques.
Concatenation of the pulse sequence can reduce a systematic error to an arbitrarily small level.
\end{abstract}

\maketitle

\def\U   {U}
\def\nU  {\bar{U}}
\def\Ro  {R_0}
\def\Rt  {R_t}
\def\nRo {\bar{R}_0}
\def\nRt {\bar{R}_t}
\def\nRto {\tilde{R}_0}
\def\V   {V}

\section{Introduction}

Quantum error correction is perhaps the biggest hurdle in building a quantum computer.
Imperfect control operations are one of several sources of error.
While error-correction schemes designed to correct for general errors \cite{Preskill97} 
no doubt also
correct control errors, error-correction or error-avoidance schemes tuned to the dominant physical error model are more efficient and practical.  Specialized error-correction schemes can also tolerate higher noise rates.

This paper specializes to systematic control errors of the following form.  When we try to apply the single-qubit pulse $\U = \exp(i \theta \hat{n} \cdot \vec{\sigma})$, a $2\theta$ rotation about axis $\hat{n}$, we in fact apply $\nU = \U \V$.  The error is systematic in the sense that it is invertible; attempting to apply $\U^\dagger$ in fact applies $\nU^\dagger$.  The form of the error $\V$ is of course restricted.  Previous authors \cite{McHughTwamley04,BrownHarrowChuang04} have considered the case of linear over-rotational errors: $\V = \exp(i \epsilon \theta \hat{n} \cdot \vec{\sigma})$ where $\epsilon$ is fixed and small, but unknown.  Here, we consider the case of general over-rotational errors, $\V = \exp[i \epsilon(\abs{\theta},\hat{n}) \hat{n} \cdot \vec{\sigma}]$.  The amount of over-rotation, $\epsilon(\abs{\theta},\hat{n})$, can now depend arbitrarily on the rotation angle $2\theta$ and also the axis of rotation $\hat{n}$.


Our error-correction method is a new composite pulse sequence, inspired by the generalization of quantum search known as amplitude amplification \cite{BrassardHoyerTapp98}.  In this algorithm, the amplitude produced in a particular target subspace by applying some unitary $\nU$ to a source state is amplified by successively repeating $R_0(\pi) \nU^\dagger R_t(\pi) \nU$.  Here $\Ro(\pi)$ and $\Rt(\pi)$ are selective reflections about the source and target, respectively.
In standard quantum search, the source is $\ket{0^n}$, $\nU = H^{\otimes n}$ transverse Hadamard, and the target is a bit string $\ket{x}$.  In the subspace spanned by $\nU \ket{0^n}$ and $\ket{x}$, the state vector steadily rotates toward $\ket{x}$.
Eventually, it rotates past the target.  

\begin{figure}[b]
\begin{center}
\includegraphics*[bb=147 564 318 645]{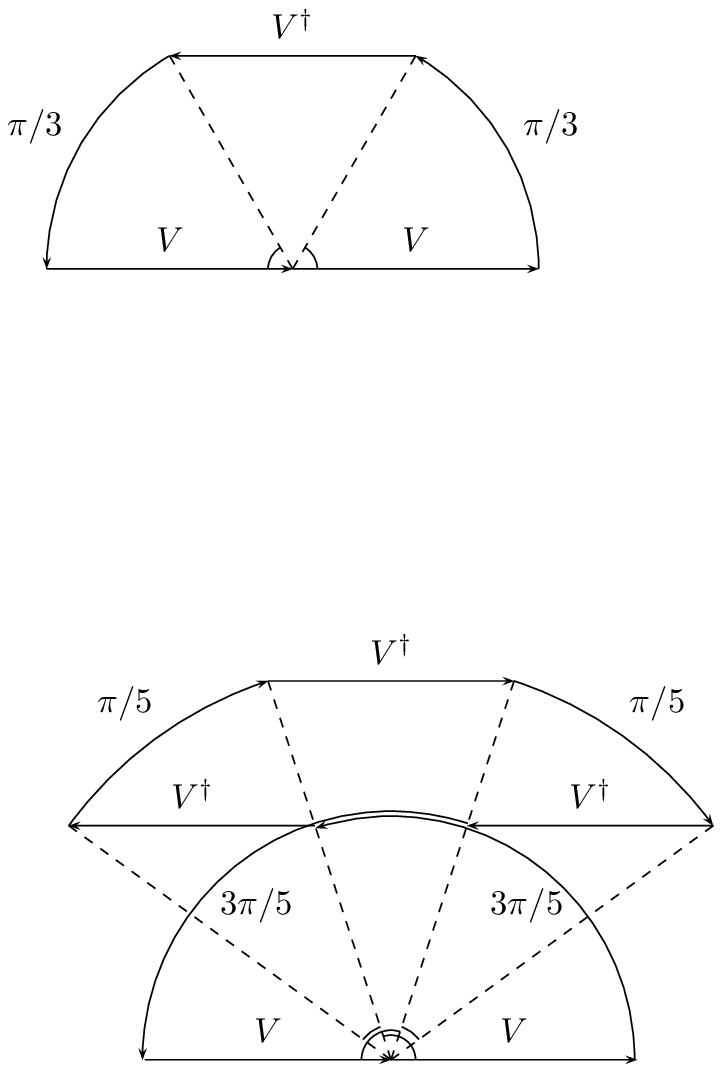}\\
\vspace{.3cm}
\includegraphics*[bb=166 338 356 465]{anglesintuition}
\end{center}
\caption{Looking down along the vector $\ket{s}$, the top diagram shows why 
when $\bra{t} \nU \ket{s} = \bra{s} V \ket{s}$ is close to one,  
$\bra{t} \nU \Ro(\pi/3) \nU^\dagger \Rt(\pi/3) \nU \ket{s} = \bra{s} \V \Ro \V^\dagger \Ro \V \ket{s}$ is even closer to one.  The bottom diagram gives similar intuition for a longer pulse sequence.} \label{f:anglesintuition}
\end{figure}

What happens in quantum search if we don't merely reflect about the source and target, but instead add a phase other than $\pi$?  It is well known that any phase bounded away from 0 works to give a square-root speedup (with different constants); for example, this fact is used in a stronger form in Ambainis's element distinctness algorithm \cite{Ambainis03}.  
One of us (LG) noticed that concatenating the basic sequence
$$
\nU \Ro(\pi/3) \nU^\dagger \Rt(\pi/3) \nU 
$$
results in the state converging to the target subspace and not overshooting it, when viewed at times $3^k$, $k \in \bf{N}$ \cite{Grover05}.  Figure~\ref{f:anglesintuition} gives some geometrical intuition.

For the present problem of systematic control errors, there is no source or target -- 
we desire a ``fully compensating" pulse sequence accurate on an arbitrary input
-- but a similar calculation still applies.  We need merely choose a source arbitrarily, say the $X$ $+1$ eigenstate $\ket{+}$, and set the target accordingly (to $\U \ket{+}$).  
Assume $\hat{n} = (0,0,1)$.  Let $\Ro = \exp(i \tfrac{\pi}{6} X)$ a $\pi/3$ rotation about the $X$ axis, and $\Rt = \U \Ro \U^\dagger$.  When we apply the sequence of noisy pulses
$$
\nRt^\dagger \nU \nRo \nU^\dagger \nRt \nU \nRo^\dagger ,
$$
the different systematic errors in both $\nU$ and the noisy $\pi/3$ rotations largely cancel out, leaving behind a higher-order error.  
(The extra pulses at either end adjust for the phase difference between $\ket{+}$ and its orthogonal complement $\ket{-}$ which would otherwise be introduced.)
This correction sequence can be concatenated on itself in a certain way to reduce errors arbitrarily.  A directly related method also applies to over-rotational errors in two-qubit gates \cite{JonesPLA03}.
Therefore this composite pulse sequence allows for an arbitrarily accurate set of universal gates, giving a threshold result for this error model.

We also consider another error model 
of systematic errors in even the rotation axis $\hat{n}$: $\V = \exp(i \vec{\epsilon} \cdot \vec{\sigma})$.  Here the error $\vec{\epsilon}$ may depend on the rotation angle $2\theta$ but, except for a specific coordinate change, not on the axis $\hat{n}$.
For example, $\Ro$ and $\Rt$ are related by the coordinate change $\Rt = \U \Ro \U^\dagger$.  We require that the errors 
be related by the same coordinate change, or $\nRt = \U \nRo \U^\dagger$.

Section~\ref{s:perfectpulses} describes the basic idea behind the composite pulse sequence, by explaining its behavior when the $\pi/3$ rotations are perfect.  In the two following sections, we extend the error model to the two cases described above.

Composite pulse sequences are an important, practical quantum control tool for removing systematic errors in a variety of quantum information processing implementations \cite{CumminsJones00, Riebeetal04}.
We need to show that our correction sequence remains practical.
While the error models we address are more general than the linear over-rotational errors which have previously been considered, the control requirement is also stricter.  We typically require the ability to rotate about an arbitrary axis in the Bloch sphere, not just one in the xy plane.
If rotations are only allowed about axes in the xy plane as in most NMR-type models, then our method only applies to correct $\pi$ pulses.
In Sec.~\ref{s:nmr}, we compare our method, with $\pi$ pulses and linear over-rotational errors, to previous fully-compensating composite pulse sequences, particularly those recently discovered in \cite{BrownHarrowChuang04}.

\section{Perfect $\pi/3$ pulses} \label{s:perfectpulses}

It is instructive to start with just the core idea of our composite pulse sequence, and build up the analysis from there.  Consider the sequence
\beq\label{e:compseq}
\nU^{(X)} = \nRt^\dagger \nU \nRo \nU^\dagger \nRt \nU \nRo^\dagger = \U \nRto^\dagger \V \nRo \V^\dagger \nRto \V \Ro^\dagger ,
\eeq
where again $\V \equiv \U^\dagger \nU$.  
Here we maintain a distinction between $\nRto \equiv \U^\dagger \nRt \U$ and $\nRo$ because the errors in the two terms might be different.  For the rest of this section, however, assume $\nRo$ and $\nRt$ are perfect, so $\nRto = \nRo = \Ro \equiv \exp(i \tfrac{\pi}{6} X)$.


Write $\V = 
\exp(i \vec{\epsilon} \cdot \vec{\sigma})$, where $\sigma = (X, Y, Z)$.
Generally, each term of $\vec{\epsilon}$ may be nonzero.
To measure the closeness of $\nU^{(X)}$ to $\U$, we compute a power series expansion of $\boldsymbol{(}\!\tr (X \cdot \U^\dagger \nU^{(X)}), \tr (Y \cdot \U^\dagger \nU^{(X)}), \tr (Z \cdot \U^\dagger \nU^{(X)}) \boldsymbol{)}$.  We obtain
\begin{multline*}
\boldsymbol{(} 2 i \epsilon_x - i \sqrt{3} (\epsilon_y^2 + \epsilon_z^2) + O(\abs{\vec{\epsilon}}^3),
\\2 i \epsilon_y^3 + 2i \epsilon_y \epsilon_z^2 + O(\abs{\vec{\epsilon}}^5),
\\2 i \epsilon_z^3 + 2i \epsilon_z \epsilon_y^2 + O(\abs{\vec{\epsilon}}^5)\boldsymbol{)} .
\end{multline*}
The first term is first order in $\epsilon_x$ because our correction $R_0$ is a rotation about the $x$ axis, and commutes with errors in the $X$ direction.  Errors in the $Y$ and $Z$ directions are symmetrically cancelled out, leaving only third-order terms.

We can express this result quite simply.
Assume $\epsilon_{x,y,z}$ is an $a,b,c$th order term.
Then the $X$ direction error order after the $X$ direction composite pulse correction is applied is $\min \{a, 2b, 2c\}$, the $Y$ direction error order is $\min\{3b, b+2c\}$ and symmetrically for the $Z$ error.
In shorthand, we write
\beq\label{e:perfectrot}
(a;b;c) \underset{X}{\rightarrow} (a,2b,2c ; 3b, b+2c ; 3c, c+2b ) .
\eeq
The underset
$X$ here refers to $X$ correction, and it is understood that we take a minimum on each of the three terms on the right.  This notation lets us quickly understand what happens when we
concatenate
correction sequences.  To concatenate when $\pi/3$ pulses are perfect, just substitute the previous level's composite pulse sequence for $\nU$.  A level $k$ concatenation will require $n_k = 3 n_{k-1} + 4$ pulses, so the sequence length grows like $4^k$.
For example, starting with only first-order $Z$ error, and applying an $X$ correction gives
$$
(\infty ; \infty ; 1) \underset{X}{\rightarrow} (2 ; \infty ; 3) .
$$
At this point, it is best to apply a $Y$ correction, since that cancels out errors in both $X$ and $Z$ directions.  ($Y$ correction is symmetrical to $X$ correction, except with $\pi/3$ rotations about the $y$ axis and the same axis conjugated by $\U$.)  At the next level of concatenation, $Z$ correction will be optimal, and so on:
\begin{eqnarray} 
(\infty ; \infty ; 1)
&\underset{X}{\rightarrow}& (2 ; \infty ; 3) \underset{Y}{\rightarrow} (6 ; 4 ; 7) \nonumber\\
&\underset{Z}{\rightarrow}& (14 ; 12 ; 7) \underset{X}{\rightarrow} (14 ; 26 ; 21) \nonumber\\
&\underset{Y}{\rightarrow}& (42 ; 26 ; 49) \underset{Z}{\rightarrow} (94 ; 78 ; 49) .
\end{eqnarray}
After $3^6 = 729$ pulses of $\nU$ or $\nU^\dagger$, and $1456$ perfect $\pi/3$ pulses (about six axes), the error is only $O(\abs{\vec{\epsilon}}^{49})$.


\begin{remark}[Generalization]
The question of whether this pulse sequence generalizes deserves further study.  We have been able to find a pulse sequence with five applications of $\nU$ or $\nU^\dagger$, and perfect rotations by $\pi/5$ or $3\pi/5$, which on input $\ket{\pm} \equiv \tfrac{1}{\sqrt{2}}(\ket{0}\pm\ket{1})$ achieves a fidelity error of $O(\epsilon^{10})$:
\begin{multline} 
\abs{\bra{\pm} \U^\dagger \nU R_0(\tfrac{3\pi}{5}) \nU^\dagger R_t(-\tfrac{\pi}{5}) \nU R_0(-\tfrac{\pi}{5}) \nU^\dagger R_t(\tfrac{3\pi}{5}) \nU \ket{\pm}}^2 \\= 1- O(\epsilon^{10}) .
\end{multline}
(See Fig.~\ref{f:anglesintuition} for geometrical intuition.)  
However, this sequence gives no improvement with imperfect correction rotations.


\end{remark}

\begin{remark}[Error measurement]
For us it is key to measure the direction of the error, as well as its magnitude.  
How does our method of measuring error compare to other reasonable methods?
On a particular input state, the difference in the fidelity from one is quadratically smaller than our measure.
The so-called infidelity between $\U$ and $\nU^\star$, or $1-\tfrac{1}{2}\abs{\tr \U^\dagger \nU^\star}$ is used in \cite{McHughTwamley04,CumminsJones00,JonesPLA03},
and is also quadratically smaller than our measure.
Brown et al \cite{BrownHarrowChuang04} use as their measure of distance the trace distance $\tr\abs{\U-\nU^\star}$, which depends on the global phase of the operators.  Our correction sequence does not give higher order accuracy in the global phase, but a simple modification of the trace distance optimizes over global phases, and then this measure of error is of the same order as ours.
\end{remark}


\section{Imperfect $\pi/3$ pulses, error angle-dependent \& axis-independent}

Let us consider the more realistic case that the $\pi/3$ pulses are themselves erroneous.
Assume that the error in a rotation depends only on the rotation angle, not on the rotation axis.  When we attempt to apply $\exp(i \theta \hat{n} \cdot \vec{\sigma})$, we actually apply $\exp\{i [\theta + \epsilon(\theta)] \hat{n} \cdot \vec{\sigma}\}$.  Here $\epsilon(\theta)$ can be an arbitrary function of $\theta$, which is however always small [order $O(\abs{\epsilon})$].  The error amount does not depend on the rotation axis $\hat{n}$.  Previous work has only considered the less-general case of linear errors, $\epsilon(\theta) = \epsilon \theta$.

In fact, 
let us generalize our calculations slightly further.  We will allow errors in the $\pi/3$ pulses besides just over-rotation errors, except these errors must be the same under change of coordinates by $\U$.  In particular, $\nRt = \U \nRo \U^\dagger$.
It isn't clear in what physical models this will be an appropriate base error assumption -- perhaps one in which the entire apparatus for applying a $\pi/3$ rotation is rotated about the qubit in three dimensions as $\U$ acts in $\bf{R^3}$, or equivalently, the qubit is physically rotated.
However, the added generality will be necessary for considering concatenation of this correction sequence.

Write $\nRo = \Ro \exp(i \vec{\delta} \cdot \vec{\sigma})$.  We obtain
\begin{widetext}
\begin{eqnarray*}
\tr (X \cdot \U^\dagger \nU^{(X)})
&=&
2 i \epsilon_x + 2 i (\sqrt{3} \delta_y + \delta_z) \epsilon_y - i \sqrt{3} (\epsilon_y^2 + \epsilon_z^2) - 2 i (\delta_y - \sqrt{3} \delta_z) \epsilon_z
+ O(\abs{\vec{\epsilon}}^3 + \abs{\vec{\delta}} \abs{\vec{\epsilon}}^2 + \abs{\vec{\delta}}^2 \abs{\vec{\epsilon}} )
\\
\tr (Y \cdot \U^\dagger \nU^{(X)})
&=&
\left(
\begin{split}2 i (\sqrt{3} \delta_y + \delta_z) \epsilon_x - 4 i (\sqrt{3} \delta_x - \sqrt{3} \delta_y \delta_z + \delta_z^2) \epsilon_y - \sqrt{3} i (\sqrt{3} \delta_y + \delta_z) \epsilon_y^2 + 2 i \epsilon_y^3 \\ - 2 i (\sqrt{3} \delta_y^2 - 2 \delta_y \delta_z - \sqrt{3} \delta_z^2) \epsilon_z - \sqrt{3} i (\sqrt{3} \delta_y - \delta_z) \epsilon_z^2 + 2 i \epsilon_y \epsilon_z^2
+ O(\abs{\vec{\epsilon}}^3 + \abs{\vec{\delta}}^2 \abs{\vec{\epsilon}} )\end{split}
\right) .
\end{eqnarray*}
$\tr (Z \cdot \U^\dagger \nU^{(X)})$ can be determined by symmetry.
In our shorthand notation, with $\delta_{x,y,z}$ being $d,e,f$th order, respectively,
\begin{equation}
(a;b;c) \underset{X}{\rightarrow} 
\left(\begin{matrix}a,e+b,f+b,2b,e+c,f+c,2c ;\\
e+a, f+a, d+b, e+f+b, 2f+b, e+2b, f+2b, 3b, 2e+c, e+f+c, 2f+c, e+2c, f+2c, b+2c ;\\
e+a, f+a, 2e+b, e+f+b, 2f+b, e+2b, f+2b, d+c, 2e+c, e+f+c, 2b+c, e+2c, f+2c, 3c \end{matrix}\right)  .
\end{equation}
\end{widetext}
For example, taking $d = e = f = \infty$, we recover Eq.~\eqref{e:perfectrot} from the perfect $\pi/3$ pulse case.  In the case of first-order over-rotation, $d=1$, $e=f=\infty$,
\begin{equation} 
(\infty; \infty; 1) 
\underset{X,Y,Y}{\longrightarrow}
(4 ; 4 ; 4)  .
\end{equation}
To obtain arbitrarily accurate rotations, it is most effective to correct both the applications of $\nU$ and the $\pi/3$ correction pulses.  So at this point correct the $\pi/3$ pulses until $d = e = f = 4$.  Note that applying such a correction maintains the invariant that the error in a pulse depend only on the angle and not the axis.  Now
\begin{equation} 
(4 ; 4 ; 4) 
\underset{X,Y,Z}{\longrightarrow} 
(12 ; 12 ; 12)  .
\end{equation}
Every three levels of concatenation (both on $\nU$ and the $\pi/3$ pulses) increases the error order by a factor of three.
Therefore obtaining error tolerance to a desired amount $\epsilon_\star$ requires poly-logarithmically many pulses in $1/\epsilon_\star$.

\section{Imperfect $\pi/3$ pulses, error both angle- and axis-dependent}

What if the error in $\pi/3$ pulses depends on which basis they are carried out in, i.e., $\nRt \neq \U \nRo \U^\dagger$?  Can we still obtain arbitrarily accurate pulses?  Perhaps surprisingly, the answer is yes, if the error is of a restricted form: only over-rotation errors.  However, the orders will not grow exponentially quickly in the number of concatenation levels, only linearly, implying that error tolerance to an amount $\epsilon_\star$ will require polynomially many pulses in $1/\epsilon_\star$ instead of only poly-logarithmically many.

Write $\nRo = \Ro \exp(i \delta X)$, $\nRt = \U \Ro \exp(i \hat{\delta} X) \U^\dagger$.  Expanding $\boldsymbol{(}\!\tr (X \cdot \U^\dagger \nU^{(X)}), \tr (Y \cdot \U^\dagger \nU^{(X)}), \tr (Z \cdot \U^\dagger \nU^{(X)})\!\boldsymbol{)}$, we obtain
$$
\left(\begin{smallmatrix}
2 i \epsilon_x - i \sqrt{3} (\epsilon_y^2 + \epsilon_z^2) + O(\abs{\delta}\epsilon^2 + \abs{\vec{\epsilon}}^3) ,\\
-2 i \sqrt{3} (\delta + \hat{\delta}) \epsilon_y - 2 i (\delta - \hat{\delta}) \epsilon_z + 2 i \epsilon_y^3 + 2 i \epsilon_y \epsilon_z^2 + O(\delta^2 \abs{\vec{\epsilon}} + \abs{\vec{\epsilon}}^5) ,\\
-2 i \sqrt{3} (\delta + \hat{\delta}) \epsilon_z - 2 i (\delta - \hat{\delta}) \epsilon_y + 2 i \epsilon_z^3 + 2 i \epsilon_z \epsilon_y^2 + O(\delta^2 \abs{\vec{\epsilon}} + \abs{\vec{\epsilon}}^5)
\end{smallmatrix}\right) ,
$$
assuming $\abs{\hat{\delta}} = \Theta(\abs{\delta})$.
In our shorthand notation, with $\delta$ first order, the rule is
\begin{multline}
(a;b;c) \underset{X}{\rightarrow} \\ (a,2b,2c ; 3b, b+2c, 1+b, 1+c ; 3c, c+2b, 1+b, 1+c )  .
\end{multline}
For example,
\begin{eqnarray} 
(\infty ; \infty ; 1)
&\underset{X}{\rightarrow}& (2 ; 2 ; 2) \underset{X}{\rightarrow} (2 ; 3 ; 3) \nonumber\\
&\underset{Y}{\rightarrow}& (3 ; 3 ; 3) \underset{X}{\rightarrow} (3 ; 4 ; 4)  ,
\end{eqnarray}
and so on, with every two levels of concatenation increasing the error order by one.  Note that we do not concatenate corrections onto the $\pi/3$ pulses, because then the error would no longer be simply over-rotational.  (Even with a more general expansion,
it turns out that the convergence is still only be linear in the number of concatenation levels.)

\section{$\pi$ pulses in NMR}\label{s:nmr}

While our method corrects against more general types of errors than previous composite pulse sequences, it also has a stronger requirement.  Namely, we must be able to apply a $\pi/3$ rotation about the $x$, $y$ and $z$ axes, and also about those same axes in the $\U$-transformed basis.  In most current proposed quantum information implementations, primitive rotations are only allowed about axes in the $xy$ plane.
For $\U$, $\Rt$ and $\Ro$ all to be rotations about axes in the $xy$ plane, it must be that $\U$ is a rotation by an integer multiple of $\pi$.  This is a considerable restriction on the applicability of our method.  Still, for $\pi$ pulses our correction succeeds in a setting more general than that for which previous methods could correct; for example, we can correct for non-linear over-rotations.

Assume now that the systematic error is in fact a linear over-rotation; when we try to apply $\exp(i \theta \hat{n} \cdot \vec{\sigma})$, we actually apply $\exp[i \theta (1+\epsilon) \hat{n} \cdot \vec{\sigma}]$.
Assume $\U = \exp(i \tfrac{\pi}{2} X) = X$.
Our $\pi/3$ correction method, correcting in the $Y$ direction, leaves behind second-order errors, with a sequence length of $3\pi + 4 \pi/3$.  (Practical composite pulse sequences need to be as short as possible, in order to minimize any non-ideal effects not accounted for in our error model.)  We cannot concatenate a $Z$ correction, but can concatenate alternately $X$ and $Y$ corrections to reduce the error arbitrarily.  Concatenating an $X$ correction onto the $Y$ correction leaves behind a third-order error, with a sequence length of $(14\!\tfrac{1}{3}) \pi$.

How does the $\pi/3$ correction method compare with 
previous correction methods?  The most practical correction methods previously known were the B2 (also known as BB1) correction sequence, and the recently discovered B4 sequence \cite{BrownHarrowChuang04}.  
These sequences are implemented as follows:
\begin{eqnarray}
\text{B2:} & \left(\bar{R}_{\phi_1}(\pi)\bar{R}_{3\phi_1}(2\pi)\bar{R}_{\phi_1}(\pi)\right) \nU \\
\text{B4:}   & \left(\begin{split}(\bar{R}_{\phi_2}(\pi)\bar{R}_{3\phi_2}(2\pi)\bar{R}_{\phi_2}(\pi))^4 \\
\times (\bar{R}_{\phi_2}(-2\pi)\bar{R}_{-\phi_2}(-4\pi)\bar{R}_{\phi_2}(-2\pi)) \\
\times (\bar{R}_{\phi_2}(\pi)\bar{R}_{3\phi_2}(2\pi)\bar{R}_{\phi_2}(\pi))^4 \end{split}\right) \nU ,
\end{eqnarray}
where $\bar{R}_{\phi}(\theta) \equiv \exp( i \tfrac{\theta(1+\epsilon)}{2} (\cos \phi X + \sin \phi Y) )$, $\cos\phi_1 = -\tfrac{\theta}{4\pi}$, and $\cos\phi_2 = -\tfrac{\theta}{24\pi}$.
They leave behind third- and fifth-order errors, respectively.  The total sequence length for correcting a $\pi$ rotation is $5\pi$ for B2 and $41\pi$ for B4.  Hence our method, in this linear over-rotation error model, seems to offer little over the plain B2 sequence.  Table~\ref{f:infidelities} gives values for the infidelities $1-\tfrac{1}{2}\abs{\tr \U^\dagger \nU^{(*)}}$ for various correction sequences.  (As previously remarked, the infidelity is quadratically smaller than the trace distance, so the infidelity error orders for the three possibilities $\pi/3$ $Y$, B2 and B4 are $4$, $6$ and $10$, respectively.)

\begin{figure}
\includegraphics*[bb=92.375 3.1875 452.375 225.668,scale=.65]{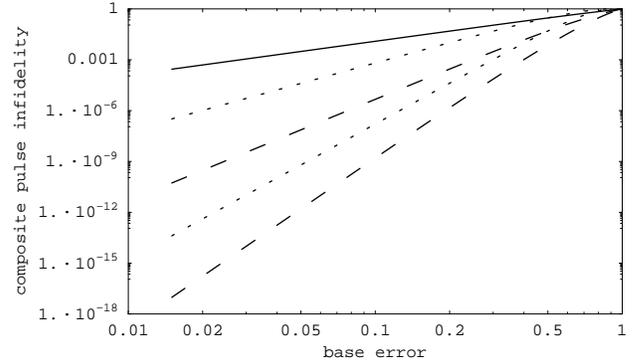}
\caption{Infidelity plotted against the base linear over-rotation parameter $\epsilon$.  Solid line: unprotected $\pi$ pulse.  Dashed line: B2 and B4 pulse sequences.  Dotted line: $\pi/3$ $Y$ correction on the $\pi$ pulse, and concatenated onto a symmetrized B2 (BB1) pulse sequence.} \label{f:fidelitygraph}
\end{figure}

Does our method complement previous correction methods?  To answer this question, we must determine the direction of the error left behind after a correction sequence.  For example, B2 and B4 are each implemented as $Bi \nU$, where $Bi$ is some particular pulse sequence not involving $\nU$.  A simple calculation shows that both B2 and B4 leave behind an error which is has relatively large $X$ and $Y$ components.  Therefore, concatenating on $X$ or $Y$ correction will not increase the error order.  We can however find an axis in the $xy$ plane which is approximately orthogonal to the $xy$ component of the error, and correct along this axis.  Alternatively, we can symmetrize the B2 and B4 sequences into $\exp[i \tfrac{\pi}{4}(1+\epsilon)X] Bi \exp[i \tfrac{\pi}{4}(1+\epsilon)X]$.  In this more symmetrical form, the error magnitude is unchanged, but the direction is entirely into the $xz$ plane.  Therefore, we can simply apply a $Y$ correction to the symmetrized sequences.  Table~\ref{f:infidelities} compares the infidelities of $Y$ correction concatenated onto the symmetrized B2 and B4 correction sequences.  Note that the former case gives fourth-order protection with a sequence length of only $(16\!\tfrac{1}{3})\pi$; this gives a new, perhaps practical, compromise between the B2 and B4 correction sequences.  Figure~\ref{f:fidelitygraph} plots the fidelities for $\epsilon > 0$.

\section{Conclusion}\label{s:conclusion}

We have presented two main results.
The $\pi/3$ correction sequence protects against general errors which depend arbitrarily on the rotation angle but not the rotation axis.
The same sequence protects against over-rotational error which depends arbitrarily on both the rotation axis and the angle of rotation.
Previously, composite pulse correction sequences 
were only known for the cases when the error was independent of the rotation axis, and depended linearly on the rotation angle.

Moreover, our composite pulse correction sequence concatenates nicely to reduce errors arbitrarily.  In the first case, the overhead number of pulses is poly-logarithmic in the desired accuracy, and in the second case the overhead is polynomial.

However, our correction sequence in general requires primitive rotations about arbitrary axes in the Bloch sphere, and only applies to correct $\pi$ pulses in the typical situation in which rotations are only allowed about axes in the xy plane.  For correcting $\pi$ pulses, our method concatenated on top of a B2 pulse correction provides a new compromise between B2 and B4.



B. R. acknowledges support from NSF ITR Grant CCR-0121555, and ARO Grant DAAD 19-03-1-0082.


\begin{table*} 
\caption{Infidelities $1-\tfrac{1}{2}\abs{\tr U^\dagger \bar{U}^{(*)}}$ of naive and variously compensated 
$\exp(i \tfrac{\pi}{2} X)$ pulses.}\label{f:infidelities}
\begin{tabular}{r @{.} l @{\qquad} r @{.} l @{\qquad} r @{.} l @{\qquad} r @{.} l @{\qquad} r @{.} l @{\qquad} r @{.} l @{\qquad} r @{.} l}
\hline \hline
\multicolumn{2}{l}{$\epsilon$} & \multicolumn{2}{l}{naive} & \multicolumn{2}{l}{B2 (BB1)} & \multicolumn{2}{l}{B4} & \multicolumn{2}{l}{$\pi/3$ $Y$} & \multicolumn{2}{l}{$\pi/3$ $Y$ $\circ$ B2} & \multicolumn{2}{l}{$\pi/3$ $Y$ $\circ$ B4} \\
\hline
0&3   & 1&1 $\times 10^{-1}$ & 3&0 $\times 10^{-3}$ & 7&2 $\times 10^{-5}$ & 4&9 $\times 10^{-2}$ & 1&0 $\times 10^{-3}$ & 2&4 $\times 10^{-5}$ \\
0&1   & 1&2 $\times 10^{-2}$ & 4&6 $\times 10^{-6}$ & 1&6 $\times 10^{-9}$ & 6&5 $\times 10^{-4}$ & 1&6 $\times 10^{-7}$ & 5&6 $\times 10^{-11}$ \\
0&03  & 1&1 $\times 10^{-3}$ & 3&4 $\times 10^{-9}$ & 9&7 $\times 10^{-15}$ & 5&2 $\times 10^{-6}$ & 1&0 $\times 10^{-11}$ & 2&9 $\times 10^{-17}$ \\
0&01  & 1&2 $\times 10^{-4}$ & 4&7 $\times 10^{-12}$ & 1&7 $\times 10^{-19}$ & 6&4 $\times 10^{-8}$ & 1&6 $\times 10^{-15}$ & 5&5 $\times 10^{-23}$ \\
0&003 & 1&1 $\times 10^{-5}$ & 3&4 $\times 10^{-15}$ & 9&8 $\times 10^{-25}$ & 5&1 $\times 10^{-10}$ & 1&0 $\times 10^{-19}$ & 2&9 $\times 10^{-29}$ \\
0&001 & 1&2 $\times 10^{-6}$ & 4&7 $\times 10^{-18}$ & 1&7 $\times 10^{-29}$ & 6&3 $\times 10^{-12}$ & 1&5 $\times 10^{-23}$ & 5&5 $\times 10^{-35}$ \\
\hline \hline
\end{tabular}
\end{table*}



\end{document}